\documentclass[aps,prd,eqsecnum,12pt,showpacs,preprintnumbers,nofootinbib,%
superscriptaddress]{revtex4}
\usepackage{lscape}
\usepackage{indentfirst}
\usepackage{latexsym}
\usepackage{multirow}

\usepackage{amssymb,amsmath}
\usepackage{graphicx}
\newcommand{\bea}{\begin{eqnarray}}
\newcommand{\eea}{\end{eqnarray}}
\newcommand{\be}{\begin{equation}}
\newcommand{\ee}{\end{equation}}
\newcommand{\bes}{\begin{subequations}}
\newcommand{\ees}{\end{subequations}}
\newcommand{\bfr}{\mathbf{r}}
\def\nn{\nonumber \\}

\def\nn{\nonumber \\}

\begin{document}

\title{Exact noise kernel for quantum fields in static de Sitter and conformally-flat spacetimes}

\author{Jason D. Bates} \email{batej6@gmail.com} \affiliation{Department of Physics, Tamkang University, Tamsui, New Taipei City,
Taiwan}

\author{Hing-Tong Cho} \email{htcho@mail.tku.edu.tw} \affiliation{Department of Physics, Tamkang University, Tamsui, New Taipei City,
Taiwan}

\author{Paul R. Anderson} \email{anderson@wfu.edu} \affiliation{Department of Physics, Wake Forest University,
Winston-Salem, North Carolina  27109, USA}

\author{B. L. Hu} \email{blhu@umd.edu} \affiliation{Maryland Center for Fundamental Physics, University of Maryland, College Park, Maryland 20742-4111, USA}

\date{Jan 8, 2013}

\begin{abstract}

We compute exact expressions of the noise kernel, defined as the expectation value of the symmetrized connected stress energy bitensor,  for conformally-invariant scalar fields with respect to the conformal vacuum, valid for an arbitrary separation (timelike, spacelike and null) of points in a class of conformally-flat spacetimes.   We derive explicit expressions for the noise kernel evaluated in the static de Sitter coordinates with respect to the Gibbons-Hawking vacuum and analyze the behavior of the noise kernel in the region near the cosmological horizon.  We develop a quasi-local expansion near the cosmological horizon and compare it with the exact results.  This gives insight into the likely range of validity of the quasi-local approximation expressions for the noise kernel for the conformally invariant scalar field in Schwarzschild spacetime which are given in \cite{EBRAH}.

\end{abstract}


\maketitle

\section{Introduction}
\label{dSintro}

While studies of quantum field theory in curved space \cite{bd-book} have yielded fundamental results such as black hole evaporation~\cite{Haw74}, and semiclassical gravity provided the framework for inflationary cosmology \cite{Guth} and understanding the origin of structures in the early universe~\cite{inflation-review}, the necessity of including the effects of fluctuations and correlations of the stress-energy tensor for quantized fields is becoming increasingly noteworthy. These aspects enter in an essential way in establishing the criteria for the validity of semiclassical gravity~\cite{HuRouVer07,And-Mol-Mot-1}. They also play an important role in cosmology, due to anticipated new data from planned precision cosmological observation experiments, and in the study of quantum effects in black hole spacetimes.  To date, investigations with emphasis on these aspects have been made of both density  perturbations during inflation~\cite{RouVer,PRV,WNF07,FMNWW} and fluctuations of quantum matter fields near a black hole and their backreaction~\cite{HuRou07} on the evolution of  black holes which may shed some light on the end-state of black hole evaporation. 

A consistent theory which captures the matter field fluctuations and the induced metric fluctuations is stochastic gravity~\cite{HVCQG,stograLivRev}.  While semiclassical gravity is based on the semiclassical Einstein equation which has as its source the vacuum expectation values of the stress energy tensor of quantum fields, stochastic gravity has at its heart the Einstein-Langevin equation from which one can study the features and dynamics of induced metric fluctuations (sometimes called `spacetime foam').  This equation is driven by the symmetrized two-point function of the stress-energy tensor of the quantum fields, called the noise kernel.

For the past decade, much of the focus of stochastic gravity has been on the calculation of the noise kernel for free scalar fields in various spacetimes.  Beginning with the derivation of a general expression for the noise kernel in arbitrary spacetimes in terms of products of derivatives of the Wightman function~\cite{martin99b, roura99b, PH01}, computations of the noise kernel have been made in Minkowski~\cite{martin99a, martin00,phillips03,HuRou07,EBRAH}, de Sitter~\cite{roura99,pn09}, anti-de Sitter~\cite{ChoHu11}, and Schwarzschild~\cite{PH03,EBRAH} spacetimes.

de Sitter space, in particular, is at the center of attention for cosmological applications due to the potential for observations related to quantum processes in the inflationary universe.  Investigations into fluctuations of the stress-energy tensors for quantized fields have been used to place constraints on the duration of inflation~\cite{WNF07} by considering their effects upon the expansion of a congruence of timelike geodesics. Additionally, such fluctuations have been found to generate non-scale invariant and non-Gaussian corrections to calculations of primordial density perturbations~\cite{FMNWW} whose signature is potentially observable in the cosmic microwave background radiation~\cite{WHFN}.

de Sitter space can also aid in the study of quantum fluctuations in black hole spacetimes.  In general the black hole backreaction and fluctuations problem~\cite{HRS} entails studying the effects upon the evolution of a black hole of both the emitted Hawking radiation and the metric fluctuations driven by the fluctuations of the quantum field.  A program for such investigation, outlined by Sinha, Raval and Hu~\cite{SRH}, is the stochastic gravity upgrade (via the Einstein-Langevin equation) of investigations carried out for the mean field in semiclassical gravity
(through the semiclassical Einstein equation) by York~\cite{york85, York} and by York and his collaborators~\cite{York2}.  However, what stalls the progress in this program is that the radial function of the Schwarzschild metric is only accessible numerically (not to mention the solution to the semiclassical Einstein equation, which is the correct background solution to use in the solution of the Einstein-Langevin equation).  An expression for the noise kernel in the near horizon region is needed for proper treatment of the backreaction and fluctuations problem.  This leads one to inquire whether some well-known approximation scheme can be applied.

In a recent paper~\cite{EBRAH},  an approximate expression for the noise kernel for the conformally invariant scalar field in Schwarzschild spacetime was derived using a quasi-local expansion.  Although we intuitively expect this approximation to be valid for separations of the points on the order of the mass scale in the region near the horizon, it is not known if the approximation remains valid when either of the points is at the horizon.  In addition, without an exact expression with which to compare, it is difficult to ascertain the exact range of point separations for which the approximation remains valid. In light of these difficulties, we would like to find an alternative way to investigate the behavior of quantum fluctuations near a horizon and to test the validity of the quasi-local approximation used there.  Fortunately, de Sitter space offers us a way to accomplish both of these tasks.

In the usual static coordinates, de Sitter space has a metric structure near the cosmological horizon that is mathematically of the same form as that of Schwarzschild spacetime.  Thus, it is expected that the noise kernel, when evaluated in the static coordinates in the region near the cosmological horizon, will have the same type of behavior as the noise kernel near the event horizon in Schwarzschild spacetime.  Because the Wightman function for de Sitter space is known, it is possible to obtain an exact, closed form expression for the noise kernel.

To compute the noise kernel in de Sitter space for the conformally invariant scalar field, we first compute an explicit expression for the noise kernel in the Minkowski vacuum state in terms of the coordinate separation.  This expression, together with the rule for the conformal transformation of the noise kernel for this field derived in Ref.~\cite{EBRAH}, provides us with a convenient means to compute exact expressions for the noise kernel of the conformal vacuum state for a large class of conformally flat spacetimes - those with metrics that are conformal to the full Minkowski space (or at least enough of it to contain a Cauchy surface).\footnote{For other conformally flat metrics a different vacuum state is appropriate. For example, de Sitter space in either the comoving or the closed coordinates is conformal to Minkowski space (with $t>0$ for the comoving coordinates), and the conformal vacuum is the Minkowski vacuum specified by some constant time Cauchy surface.  In contrast, in the usual static coordinates de Sitter space is conformal to Rindler space, and it is therefore the Rindler vacuum which is the preferred conformal vacuum state~\cite{bd-book}.  Anti-de Sitter space, in contrast, is conformal to Minkowski space with a timelike boundary (for instance, at $x=0$).  Since all Cauchy surfaces cross the boundary the choice of conformal vacuum must respect the appropriate boundary conditions~\cite{Isham78}.}

We specialize this result to de Sitter spacetime in spatially flat Robertson-Walker coordinates and then make a coordinate transformation to the usual static coordinate system.  Note that this ensures that the vacuum state is the correct one.  Alternatively one could make a direct conformal transformation from flat space to de Sitter space in the static coordinates using the Rindler vacuum state; both methods result in identical expressions for the noise kernel in the static coordinates.  The resulting expression for the noise kernel is used to investigate its behavior in the region near the cosmological horizon.  This behavior is expected to be similar to that which would be found for the exact noise kernel for the conformally invariant scalar field in a static black hole spacetime near the event horizon if the field is in the Hartle-Hawking state~\cite{hartle-hawking}.

In addition, we investigate the likely range of validity of the quasi-local expansion of the noise kernel for Schwarzschild spacetime given in~\cite{EBRAH}.  This is accomplished by computing a similar expansion for the noise kernel in static de Sitter space in terms of inverse powers of the coordinate separation and truncating the series at the same order that was done in~\cite{EBRAH} for Schwarzschild spacetime.  We compute the relative error between the exact expression and the truncated series when the coordinates are split in the time and radial directions, and investigate the range of separations for which that error remains small.

In Sec.~\ref{conf_flat}, we present the results of our computation of the noise kernel for conformally flat spacetimes by giving an exact expression for an arbitrary separation of the points (including the delta-function contributions present for null separations).  In Sec.~\ref{max_sym}, we compare the exact result presented in Sec.~\ref{conf_flat} for the cases of de Sitter and anti-de Sitter spacetimes with the expressions computed by Osborn and Shore~\cite{o-s-99} and Cho and Hu~\cite{ChoHu11} for the stress energy two-point function for the conformally invariant field.  In Sec.~\ref{sdS}, we show the corresponding expression for two components of the noise kernel in the static de Sitter coordinates when the points are separated in a timelike or spacelike direction. We investigate the behavior of the noise kernel as one or both of the points approaches the cosmological horizon.  We also expand the noise kernel in inverse powers of the coordinate separation and investigate the range of validity of this expansion.  The results of these calculations are discussed in Sec.~\ref{disc}.

The conventions used throughout are those of Misner, Thorne, and Wheeler~\cite{MTW}.  Units are chosen such that $c=\hbar=G=1$.

\section{The noise kernel in conformally flat spacetimes}
\label{conf_flat}

The noise kernel for a quantum scalar field in a Gaussian state can be expressed in terms of the Wightman function~\cite{PH01, EBRAH}.  For the conformally invariant scalar field, this expression is
%
%
%
\begin{equation}
 N_{abc'd'}  =
 {\rm Re} \left\{  \bar K_{abc'd'}
  + g_{ab}   \bar K_{c'd'}
 + g_{c'd'} \bar K'_{ab}
 + g_{ab}g_{c'd'} \bar K \right\}
\label{general-noise-kernel}
\end{equation}
with\footnote{Note that the superscript $+$ on $G^+$ has been omitted for notational simplicity.}
\bes
\label{General-Noise}
\begin{eqnarray}
9  \bar K_{abc'd'} &=&
%
4\,\left( G{}\!\,_{;}{}_{c'}{}_{b}\,G{}\!\,_{;}{}_{d'}{}_{a} +
    G{}\!\,_{;}{}_{c'}{}_{a}\,G{}\!\,_{;}{}_{d'}{}_{b} \right)
%
+ G{}\!\,_{;}{}_{c'}{}_{d'}\,G{}\!\,_{;}{}_{a}{}_{b} +
  G\,G{}\!\,_{;}{}_{a}{}_{b}{}_{c'}{}_{d'} \cr
%
&& -2\,\left( G{}\!\,_{;}{}_{b}\,G{}\!\,_{;}{}_{c'}{}_{a}{}_{d'} +
    G{}\!\,_{;}{}_{a}\,G{}\!\,_{;}{}_{c'}{}_{b}{}_{d'} +
    G{}\!\,_{;}{}_{d'}\,G{}\!\,_{;}{}_{a}{}_{b}{}_{c'} +
    G{}\!\,_{;}{}_{c'}\,G{}\!\,_{;}{}_{a}{}_{b}{}_{d'} \right)  \cr
%
&& + 2\,\left(
G{}\!\,_{;}{}_{a}\,G{}\!\,_{;}{}_{b}\,{R{}_{c'}{}_{d'}} +
    G{}\!\,_{;}{}_{c'}\,G{}\!\,_{;}{}_{d'}\,{R{}_{a}{}_{b}} \right)  \cr
%
&& - \left( G{}\!\,_{;}{}_{a}{}_{b}\,{R{}_{c'}{}_{d'}} +
  G{}\!\,_{;}{}_{c'}{}_{d'}\,{R{}_{a}{}_{b}}\right) G
%
 +{\frac{1}{2}}  {R{}_{c'}{}_{d'}}\,{R{}_{a}{}_{b}} {G^2}
\end{eqnarray}
\begin{eqnarray}
 36  \bar K'_{ab} &=&
%
8 \left(
 -  G{}\!\,_{;}{}_{p'}{}_{b}\,G{}\!\,_{;}{}^{p'}{}_{a}
 + G{}\!\,_{;}{}_{b}\,G{}\!\,_{;}{}_{p'}{}_{a}{}^{p'} +
  G{}\!\,_{;}{}_{a}\,G{}\!\,_{;}{}_{p'}{}_{b}{}^{p'}
\right)\cr &&
%
4 \left(
    G{}\!\,_{;}{}^{p'}\,G{}\!\,_{;}{}_{a}{}_{b}{}_{p'}
  - G{}\!\,_{;}{}_{p'}{}^{p'}\,G{}\!\,_{;}{}_{a}{}_{b} -
  G\,G{}\!\,_{;}{}_{a}{}_{b}{}_{p'}{}^{p'}
\right) \cr
%
&& - 2\,{R'}\,\left( 2\,G{}\!\,_{;}{}_{a}\,G{}\!\,_{;}{}_{b} -
    G\,G{}\!\,_{;}{}_{a}{}_{b} \right)  \cr
%
&&  -2\,\left( G{}\!\,_{;}{}_{p'}\,G{}\!\,_{;}{}^{p'} -
    2\,G\,G{}\!\,_{;}{}_{p'}{}^{p'} \right) \,{R{}_{a}{}_{b}}
%
 - {R'}\,{R{}_{a}{}_{b}} {G^2}
\end{eqnarray}
\begin{eqnarray}
 36 \bar K &=&
2\,G{}\!\,_{;}{}_{p'}{}_{q}\,G{}\!\,_{;}{}^{p'}{}^{q}
+ 4\,\left( G{}\!\,_{;}{}_{p'}{}^{p'}\,G{}\!\,_{;}{}_{q}{}^{q} +
    G\,G{}\!\,_{;}{}_{p}{}^{p}{}_{q'}{}^{q'} \right)  \cr
&& - 4\,\left( G{}\!\,_{;}{}_{p}\,G{}\!\,_{;}{}_{q'}{}^{p}{}^{q'} +
    G{}\!\,_{;}{}^{p'}\,G{}\!\,_{;}{}_{q}{}^{q}{}_{p'} \right)  \cr
&& + R\,G{}\!\,_{;}{}_{p'}\,G{}\!\,_{;}{}^{p'} +
  {R'}\,G{}\!\,_{;}{}_{p}\,G{}\!\,^{;}{}^{p} \cr
&& - 2\,\left( R\,G{}\!\,_{;}{}_{p'}{}^{p'} +
{R'}\,G{}\!\,_{;}{}_{p}{}^{p} \right)
     G
+  {\frac{1}{2}} R\,{R'} {G^2}  \;.
\end{eqnarray}
\label{generalnoise}
\ees
Primes on indices denote tensor indices at the point $x'$ and unprimed ones denote indices at the point $x$.  $R_{ab}$ and $R_{c'\,d'}$ are the Ricci tensor evaluated at the points $x$ and $x'$, respectively; $R$ and $R'$ are the scalar curvature evaluated at $x$ and $x'$.

As shown in~\cite{EBRAH}, under a conformal transformation of the metric,
\bea \tilde{g}_{ab}(x) = \Omega(x)^2 g_{ab}(x) \;, \label{conf-trans} \eea
the noise kernel for a conformally invariant scalar field transforms as
\bea \tilde{N}_{abc'd'}(x,x') = \Omega(x)^{-2} N_{abc'd'}(x,x') \Omega(x')^{-2} \;. \label{nk-conformal} \eea
Using this result in conjunction with Eq.~\eqref{generalnoise} evaluated for the Minkowski vacuum state, we have computed an exact expression for the noise kernel for the conformally invariant scalar field in the conformal vacuum state of a large class of conformally flat spacetimes. As previously noted, these results are useful for those metrics that are conformal to the full Minkowski metric (or at least a part of it that contains a Cauchy surface). 

The Wightman function in flat space in the Minkowski vacuum is
\bea
  G^{+}(x,x') &=& \frac{1}{8\pi^2 \sigma(x,x')} + \frac{i}{8\pi} \delta(\sigma(x,x')) \text{sgn}(t-t') \;, \label{mink-vac}
\eea
with
\be \sigma(x,x') = \frac{1}{2} \left[-(x^0 -x^{0 \, '})^2 + (\vec{x}-\vec{x}^{\,'})^2 \right] \;.  \ee
In general $\sigma$ satisfies the relationship
\be \sigma = \frac{1}{2} g_{ab} \sigma^a \sigma^b = \frac{1}{2} g_{a' b'} \sigma^{a \,'} \sigma^{b \,'} \;, \ee
with
\bea \sigma^a &\equiv& \sigma^{;a}  \;, \nonumber \\
\sigma^{a\,'} &\equiv& \sigma^{;a'}  \;.
\eea
Here primes on indices indicate that the indices are at the point $x'$.
If the points are close together then
\bea \sigma^a &=& x^a - x^{'\,a} \;, \nonumber \\
      \sigma^{a \, '} &=& - (x^a - x^{'\,a}) \, \eea
and of course in general
\bea \sigma_a &=& g_{ab} \sigma^b  \nonumber \\
     \sigma_{a'} &=& g_{a' \, b'} \sigma^{b \,'} \;. \eea

The expression for the noise kernel for the conformally invariant scalar field in the conformal vacuum state is
\bea
  N_{abc'd'}(x,x') &=& \Omega(x)^{-2} \Omega(x')^{-2} \left[ \frac{\sigma_a \sigma_b \sigma_{c'} \sigma_{d'}}{48 \pi^4 \sigma^6} + \frac{\sigma_{(a} \eta_{b)(c'} \sigma_{d')}}{24 \pi^4 \sigma^5} + \frac{4 \eta_{a(c'} \eta_{d')b} - \eta_{ab} \eta_{c'd'}}{192 \pi^4 \sigma^4} \right. \nn
                            & & \left. - \frac{ \sigma_a \sigma_b \sigma_{c'} \sigma_{d'}}{576 \pi^2} \left( 9 \epsilon''(\sigma)^2 - 8 \epsilon'(\sigma)\epsilon'''(\sigma) + \epsilon(\sigma)\epsilon''''(\sigma) \right) \right. \nn
                            & & \left. + \frac{ \sigma_a \sigma_b \eta_{c'd'} +  \eta_{ab} \sigma_{c'} \sigma_{d'}}{576 \pi^2}\left(5 \epsilon'(\sigma)\epsilon''(\sigma) - \epsilon(\sigma)\epsilon'''(\sigma) \right) \right. \nn
                            & & \left. + \frac{\sigma_{(a} \eta_{b)(c'} \sigma_{d')}}{144 \pi^2} \epsilon(\sigma)\epsilon'''(\sigma) \right. \nn
                            & & \left. - \frac{\eta_{a(c'} \eta_{d')b}}{288 \pi^2}\left( 4 \epsilon'(\sigma)^2  + \epsilon(\sigma)\epsilon''(\sigma) \right) \right. \nn
                            & & \left. + \frac{\eta_{ab} \eta_{c'd'}}{576 \pi^2}\left( \epsilon'(\sigma)^2 - \epsilon(\sigma)\epsilon''(\sigma) \right) \right] \;. \label{conf-delta}
\eea
Here $(...)$ indicates symmetrization of the indices, $\eta_{ab}$ is the Minkowski metric, $\eta_{ac'} = \text{diag}(-1, 1, 1, 1)$ is the bivector of parallel transport in Minkowski space for Cartesian coordinates, and $\epsilon(\sigma) \equiv \delta(\sigma) \text{sgn}(t-t')$, $\epsilon'(\sigma) \equiv d \delta(\sigma)/d \sigma \, \text{sgn}(t-t')$, etc.  

This result agrees with previous computations of the noise kernel in the Minkowski vacuum state~\cite{martin00, HuRou07, EBRAH} and (as we show in more detail below) with results by Osborn and Shore~\cite{o-s-99} and Cho and Hu~\cite{ChoHu11} for the noise kernel in dS and AdS spacetimes.

As would be expected for a quantity made out of Green functions, the noise kernel is finite when the points are split in a timelike or spacelike direction but it is divergent when they are split in a null direction.  However, unlike the Wightman Green function in~\eqref{mink-vac}, this singularity is not integrable because one always finds a product of two identical delta functions.  The exact meaning of this divergence is under investigation.

\section{The noise kernel for maximally symmetric spacetimes.}
\label{max_sym}

In maximally symmetric spacetimes such as Minkowski space, de Sitter space or anti-de Sitter space, bitensors like the noise kernel can be alternatively expressed in terms of the geodesic separation biscalar, $\tau(x,x')$, which in Minkowski space is given by
\bea \tau(x,x') \equiv \sqrt{2\sigma} = \sqrt{-(x^0 -x^{0 \, '})^2 + (\vec{x}-\vec{x}^{\,'})^2} \;, \eea
its derivatives $n_a \equiv \nabla_a \tau(x,x')$ and $n_{c'} \equiv \nabla_{c'} \tau(x,x')$, and the bivector of parallel transport $g_{ac'}(x,x') $~\cite{a-j-86}.  Due to the symmetries involved there are five such terms in the expression for the noise kernel,
\bea
  N_{abc'd'}(x,x') &=& C_1 n_a n_b n_{c'} n_{d'} + C_2 (g_{ab} n_{c'} n_{d'} + n_a n_b g_{c'd'}) \nn
                             & & + 4 C_3 n_{(a}g_{b)(c'}n_{d')} + 2 C_4 g_{a(c'}g_{d')b} + C_5 g_{ab}g_{c'd'} \,
\eea
where the coefficients $C_1$ through $C_5$ are functions only of the geodesic separation $\tau(x,x')$.

Using this representation, Osborn and Shore~\cite{o-s-99} computed the coefficients of the stress-energy bi-tensor for free scalar fields in Euclidean spacetimes of constant curvature.  Cho and Hu~\cite{ChoHu11} have also computed them for scalar fields with arbitrary mass and coupling in Euclideanized anti-de Sitter space. Against these exact results, for the purpose of checking the range of validity of short distance approximations, we use Eq. (2.11) to compute these coefficients which gives exact results for the Minkowski vacuum, and approximate results up to order $O(\tau^{-4})$ for de Sitter and anti-de Sitter space in the conformal vacuum state.

For the Minkowski vacuum, ignoring the delta function component, the coefficients may be read off directly from Eq.~\eqref{conf-delta}.  We find that
\bea
  C_1 &=& \frac{4}{3 \pi^4 \tau^8} \nn
  C_2 &=& 0 \nn
  C_3 &=& \frac{1}{3 \pi^4 \tau^8} \nn
  C_4 &=& \frac{1}{6 \pi^4 \tau^8} \nn
  C_5 &=& -\frac{1}{12 \pi^4 \tau^8} \, .
\eea

In de Sitter space we have computed the first two terms in an expansion which is valid when the points are close together.  The result is
\bea
  C_1 &=& \frac{4}{3 \pi^4 \tau^8} - \frac{4}{9\pi^4 \alpha^2 \tau^6} + O(\tau^{-4}) \nn
  C_2 &=& 0 \nn
  C_3 &=& \frac{1}{3 \pi^4 \tau^8} - \frac{1}{9\pi^4 \alpha^2 \tau^6} + O(\tau^{-4}) \nn
  C_4 &=& \frac{1}{6 \pi^4 \tau^8} - \frac{1}{18\pi^4 \alpha^2 \tau^6} + O(\tau^{-4}) \nn
  C_5 &=& -\frac{1}{12 \pi^4 \tau^8} + \frac{1}{36\pi^4 \alpha^2 \tau^6} + O(\tau^{-4}) \, .
\eea

In the case of anti-de Sitter space, the computation proceeds slightly differently, due to the need to specify boundary conditions at infinity and the fact that anti-de Sitter space is conformally related to only half of Minkowski space~\cite{Isham78}.  Choosing Dirichlet boundary conditions, the appropriate vacuum Wightman function for Minkowski spacetime with a boundary located at $x=0$ (and with points non-null separated) is given by
\bea
  G^{+}(x,x') &=& \frac{1}{8\pi^2 \sigma} -  \frac{1}{8\pi^2 \bar{\sigma}} \;, \label{mink-vac-boundary}
\eea
where $\bar{\sigma} = \frac{1}{2}[-(t-t')^2 + (x+x')^2 + (y-y')^2 +(z-z')^2]$.

Plugging this expression into Eq.~\eqref{generalnoise} and following the same procedure used for the de Sitter result, we again find for each coefficient the first two terms in an expansion which is valid when the points are close together.  The result is
\bea
  C_1 &=& \frac{4}{3 \pi^4 \tau^8} - \frac{10}{9\pi^4 \alpha^2 \tau^6} + O(\tau^{-4}) \nn
  C_2 &=& \frac{1}{12\pi^4 \alpha^2 \tau^6} + O(\tau^{-4}) \nn
  C_3 &=& \frac{1}{3 \pi^4 \tau^8} - \frac{7}{36\pi^4 \alpha^2 \tau^6} + O(\tau^{-4}) \nn
  C_4 &=& \frac{1}{6 \pi^4 \tau^8} - \frac{5}{72\pi^4 \alpha^2 \tau^6} + O(\tau^{-4}) \nn
  C_5 &=& -\frac{1}{12 \pi^4 \tau^8} + \frac{1}{72\pi^4 \alpha^2 \tau^6} + O(\tau^{-4}) \, .
\eea

When analytically continued to the Euclidean sector, these expressions for de Sitter and anti-de Sitter agree with those computed by Osborn and Shore~\cite{o-s-99} and Cho and Hu~\cite{ChoHu11} for the vacuum two-point function of the stress tensor in 4-D spherical and hyperbolic Euclidean spacetimes.

\section{The noise Kernel in de Sitter space}
\label{sdS}

The noise kernel for the conformally invariant field in the conformal vacuum in de Sitter space can be obtained from Eq.~\eqref{conf-delta} with the substitutions
\bea
  t &\rightarrow& -\eta \nn
  t' &\rightarrow& -\eta' \nn
  \Omega(x) &=& \frac{\alpha}{(-\eta)} \nn
  \Omega(x') &=& \frac{\alpha}{(-\eta')} \label{dSnk} \;,
\eea
where we have represented de Sitter space in terms of the comoving coordinates with metric
\bea
  ds^2 =  \frac{\alpha^2}{(-\eta)^2}  (-d\eta^2 + dx^2 + dy^2 + dz^2) \;. \label{dS-conf}
\eea

However, we are primarily interested in investigating the noise kernel in the static coordinates, for which the metric has a form similar to that of Schwarzschild spacetime.  The coordinate transformation to this system is
\bea
  x &\equiv& \frac{e^{-T/\alpha}}{\sqrt{B}}\rho\sin{\theta}\cos{\phi} \;, \nn
  y &\equiv& \frac{e^{-T/\alpha}}{\sqrt{B}}\rho\sin{\theta}\sin{\phi} \;, \nn
  z &\equiv& \frac{e^{-T/\alpha}}{\sqrt{B}}\rho\cos{\theta} \;, \nn
  -\eta &\equiv& \alpha \frac{e^{-T/\alpha}}{\sqrt{B}} \;, \label{ds-static-trans}
\eea
and the resulting line element is
\bea
    ds^2 = - B dT^2 + \frac{d\rho^2}{B} + \rho^2 d\theta^2 + \rho^2 \sin^2 \theta d\phi^2 \;, \label{dS-static}
\eea
with $B=1-\frac{\rho^2}{\alpha^2}$.  For an observer situated at the origin, $B=0$ is a cosmological horizon which marks the boundary of his observable universe.  It is this coordinate system which interests us the most, since it provides an opportunity to study the noise kernel near the horizon and compare its behavior with that found for the approximate noise kernel in Schwarzschild spacetime.

From the definitions given in Eq.~\eqref{ds-static-trans}, we transform Eq.~\eqref{conf-delta} to the static coordinates using the relation
\bea
  N_{abc'd'}(x,x') &=& \frac{\partial x^A}{\partial x^a} \frac{\partial x^B}{\partial x^b} \frac{\partial x^{C'}}{\partial x^{c'}} \frac{\partial x^{D'}}{\partial x^{d'}}  N_{ABC'D'}(x,x') \;, \label{coord-trans}
\eea
where we have used capital letters to represent indicies of the comoving coordinates and lowercase to represent indicies of the static coordinates.

To avoid coordinate singularities, we express the noise kernel in terms of an orthonormal frame at each of the two points.  We do this by introducing orthonormal basis vectors at each point which satisfy
\bea
  (e_{\Hat{a}})^c(e_{\hat{b}})_c &=& \eta_{\hat{a}\hat{b}} \;, \\
  (e^{\Hat{a}})_c(e_{\hat{a}})_d &=& g_{cd} \;.
\eea
Here $\eta_{\hat{a}\hat{b}}$ is the Minkowski metric.  The components of a vector may be written in the orthonormal basis as
\bea
  A_{\hat{a}} &=& (e_{\hat{a}})^{a} A_{a} \;.
\eea
Similarly, the noise kernel in this basis is
\bea
  N_{\hat{a}\hat{b}\hat{c'}\hat{d'}}(x,x') &=& (e_{\hat{a}})^a(e_{\hat{b}})^b(e_{\hat{c'}})^{c'}(e_{\hat{d'}})^{d'}N_{abc'd'}(x,x') \;. \label{nk-ortho-gen}
\eea
For the static de Sitter coordinates, we choose basis vectors such that
\bea
  (e_{\hat{T}})^{T} &=& \sqrt{-g^{TT}} \nn
  (e_{\hat{\rho}})^{\rho} &=& \sqrt{g^{\rho\rho}} \nn
  (e_{\hat{\theta}})^{\theta} &=& \sqrt{g^{\theta\theta}} \nn
  (e_{\hat{\phi}})^{\phi} &=& \sqrt{g^{\phi\phi}} \;. \label{orth-coords}
\eea
All other components are zero.

In general, the expressions resulting from this procedure are quite long.  Although we have computed every component for the noise kernel in the static coordinates, for the sake of brevity we present only two of them:
\bes
\bea
  N_{\hat{T}\hat{T}\hat{T'}\hat{T'}}(x,x') &=&  (BB')^{-1} N_{TTT'T'}(x,x') \nn
                                      &=& \frac{1}{12 \pi ^4 \left[\alpha ^2 \left(  \sqrt{B B'}\,\tau-2 \right)+2 \rho \rho' \cos (\gamma)\right]^6} \nonumber \\
                                      & & \nonumber \\
                                      & & \times \left\{ \alpha ^4 \left[-12  \sqrt{B B'}\,\tau+B B' \left(\tau^2 + 14 \right) \right. \right. \nonumber \\
                                      & & \left. \left. - \left(2 B +2 B' -6 \right) \left(\tau^2 - 1\right)\right] \right. \nonumber \\
                                      & & \left. +4 \alpha ^2 \rho \rho' \cos (\gamma) \left(3 \sqrt{B B'}\,\tau-2 \left(\tau^2-1\right)\right) \right. \nonumber \\
                                      & & \left. +2 \rho^2 \rho'\,^2 \left(\tau^2-1\right) \cos (2 \gamma)\right\} \label{NKstatic-TTTT-orth}
\eea
\bea
  N_{\hat{T}\hat{\rho}\hat{T'}\hat{\rho'}}(x,x') &=& N_{T \rho T' \rho'}(x,x') \nn
                                       &=& \frac{\alpha^2}{6 \pi ^4 \left[\alpha ^2 \left( \sqrt{B B'}\,\tau-2 \right)+2 \rho \rho' \cos (\gamma)\right]^6} \nonumber \\
                                      & & \nonumber \\
                                      & & \times \left\{\alpha ^2 \cos (\gamma) \left[-4 \sqrt{B B'}\,\tau+ B B' \left(\tau^2+4\right) \right. \right. \nonumber \\
                                      & & \left. \left. - \left(2 B -2B' + 2\right) \left(\tau^2-2\right)\right] \right. \nonumber \\
                                      & & \left. +\rho \rho' (\cos (2 \gamma)+3) \left( \sqrt{B B'}\,\tau-\tau^2 + 2\right)\right\}  \;, \label{NKstatic-TrTr-orth}
\eea \label{NKstatic-orth}
\hskip -1ex where
\bea
  \tau &\equiv& 2 \cosh (\Delta T/\alpha) \;, \\
  \cos \gamma &\equiv& \cos \theta \cos \theta\,' + \sin \theta \sin \theta\,' \cos (\phi - \phi') \;,
\eea
\ees
and $B' = 1-\rho'\,^2/\alpha^2$.
Note that we have neglected the delta function contributions in the above expressions.  This is due to the fact that the initial computations of the delta function contributions each contained approximately $2700$ terms and we have not yet found a means by which to simplify the expressions into displayable forms.  The expression shown in Eq.~\eqref{NKstatic-orth} is valid only for non-null separations of the points.

\subsection{Behavior near the horizon}

When either of the two points approaches the cosmological horizon, we see that $\rho \rightarrow \alpha$ and $B \rightarrow 0$ (or $\rho' \rightarrow \alpha$ and $B' \rightarrow 0$, respectively).  However, note that at the horizon $T$ is a null coordinate and is infinite.  Therefore, $\Delta T$ is an ill-defined quantity when one or both points are on the horizon.  Nonetheless, if we fix the value of $\Delta T$, then inspection of Eq.~\eqref{NKstatic-orth} shows that both of the components displayed therein remain bounded when either point is arbitrarily close to the horizon, so long as the points are spacelike or timelike separated.\footnote{We find this behavior to be true for all components of the noise kernel when expressed in the orthonormal frame.}

For example, if we take $\rho'$ to be arbitrarily close to the horizon, then
\bes
\bea
  N_{\hat{T}\hat{T}\hat{T'}\hat{T'}}(x,x') &\approx& -\frac{\tau^2-1}{384 \pi^4 \alpha^4 (\alpha -\rho \cos (\gamma ))^6} \nn
      & & \times \left[ \alpha^2 (B-3) + 4 \alpha \rho \cos (\gamma) - \rho^2 \cos (2\gamma) \right] \\
  N_{\hat{T}\hat{\rho}\hat{T'}\hat{\rho'}}(x,x') &\approx&-\frac{\tau^2-2}{384 \pi^4 \alpha^3 (\alpha -\rho \cos (\gamma ))^6} \nn
      & & \times \left[ (2B-4)\alpha \cos(\gamma)+\rho(\cos(2\gamma)+3) \right] \;.
\eea
\ees

When $\rho=\rho'$ and both points are near the horizon, we have
\bes
\bea
  N_{\hat{T}\hat{T}\hat{T'}\hat{T'}}(x,x') &\approx& \frac{\tau ^2-1}{3072 \pi ^4 \alpha ^8} \csc^8\left(\frac{\gamma}{2}\right) \;. \label{NKstatic-TTTT-horizon}
\eea
and
\bea
  N_{\hat{T}\hat{\rho}\hat{T'}\hat{\rho'}} &\approx& -\frac{\tau^2-2}{3072 \pi ^4 \alpha ^8} \csc^8\left(\frac{\gamma}{2}\right) \;. \label{NKstatic-TrTr-horizon}
\eea
\ees
Therefore, the noise kernel is bounded when either or both of the points are near the horizon so long as the separation in the $T$ coordinate is fixed and either $\gamma \neq 0$, $\rho \neq \rho'$, or both.  For $\gamma = 0$ and $\rho=\rho'$, the above expression is not bounded as the two points approach the horizon.  This is expected since, on the cosmological horizon, $T$ is a null coordinate and $\Delta T$ is ill-defined.

\subsection{Comparison with Schwarzschild spacetime}
\label{schw_comp}

In~\cite{EBRAH}, an approximate expression for the noise kernel was computed for the conformally invariant scalar field in Schwarzschild spacetime.  Explicit expressions were given for the $N^{t~t'~}_{~t~t'}$ component of noise kernel in the case that the separation was only in time or only in terms of the radial coordinate $r$.

As discussed in Sec.~\ref{dSintro}, we are interested in comparing the noise kernel obtained for Schwarzschild spacetime with that obtained for de Sitter space in the static coordinates.  We do this by generating an expansion for the noise kernel in the static de Sitter coordinates that is equivalent to the quasi-local expansion used for the Schwarzschild case.  In principle, the proper way to do this is to use a quasi-local expansion for the Wightman function in the static de Sitter coordinates and recompute the noise kernel using that expression.  However, since we are primarily interested in comparing our results with Eqs.~(4.16) and~(4.17) of~\cite{EBRAH}, we can generate equivalent approximations by expanding Eq.~\eqref{NKstatic-TTTT-orth} in powers of $1/\Delta T$, $1/\Delta \rho$, and $\eta \equiv \cos \gamma - 1$ and truncating the series at the appropriate order.

For our investigation, we consider the $N_{\hat{T}\hat{T}\hat{T'}\hat{T'}}$ component and begin by splitting in the time and radial directions.  The result is
\bes
\bea
  \left[ N_{\hat{T}\hat{T}\hat{T'}\hat{T'}} \right]_{\Delta \rho = \gamma = 0} &=& \frac{1}{4 \pi^4 \alpha^8 B^4  (\tau -2)^4} \;, \label{nk-time-ex} \\
  \left[ N_{\hat{T}\hat{T}\hat{T'}\hat{T'}} \right]_{\Delta T = \gamma = 0} &=& -\frac{B+B'-2BB'+2(\sqrt{BB'}-1)\widetilde{B}}{64 \pi^4 \alpha^8 (\sqrt{BB'} - \widetilde{B})^6} \;. \label{nk-rad-ex}
\eea
\ees
Here, $\widetilde{B} \equiv 1 - \rho \rho' / \alpha^2$.

Expanding the $N_{\hat{T}\hat{T}\hat{T'}\hat{T'}}$ component in powers of $1/\Delta T$ and $1/\Delta \rho$ and truncating the series at order $O[(x-x')^{-4}]$, we find
\bes
\bea
  [N_{\hat{T}\hat{T}\hat{T'}\hat{T'}}(x,x')]_{\text{series}} &=& \frac{1}{4 \pi ^4 B^4} \left[ \frac{1}{\Delta T^8}-\frac{1}{3 \alpha^2 \Delta T^6 }+\frac{7}{120 \alpha^4 \Delta T^4} \right] \nn
                                              & & + O[\Delta T^{-2}] \;, \label{nk-time}
\eea
\bea
  [N_{\hat{T}\hat{T}\hat{T'}\hat{T'}}(x,x')]_{\text{series}} &=& \frac{1}{\pi^4} \left[ \frac{B^4}{4 \Delta \rho^8}-\frac{(B-1) B^3}{ \rho \Delta \rho^7}+\frac{(5-6 B) B^2}{4 \alpha^2 \Delta \rho^6} \right. \nonumber \\
                                              & & \nonumber \\
                                              & & \left. +\frac{\left(2 B^2-3 B+1\right) B}{2 \alpha^2  \rho \Delta \rho^5} +\frac{8 B^2-8 B+1}{32 \alpha^4 \Delta \rho^4} \right] \nn
                                              & & + O[\Delta \rho^{-3}] \;. \label{nk-rad}
\eea
\ees

In~\cite{EBRAH}, we considered thermal states in which $\kappa \equiv T/2\pi$ was left arbitrary.  However, the natural choice of state in Schwarzschild spacetime is the Hartle-Hawking state~\cite{hartle-hawking}, for which $\kappa=1/4M$, since only in this state does the stress-energy tensor remain finite at the horizon; for all other thermal states, including the zero-temperature Boulware state~\cite{boulware}, the stress-energy tensor diverges badly at $r=2M$.  We may compare the above expressions with those of~\cite{EBRAH} by setting $\kappa=1/4M$ and taking $B \rightarrow f = 1-2M/r$ and $\alpha \rightarrow 1/\kappa$.  In general, we find that although the coefficients at each order are different, the general form of the expansion is the same.  In addition, for time separations we find that to leading order in powers of $1/f$ the two expressions are identical.  For radial separations, the situation is slightly different; in the near horizon limit the only surviving term is proportional to $\kappa^4/\Delta r^4$ for Schwarzschild and $1/\alpha^4 \Delta \rho^4$ for de Sitter.  Thus, the expressions are equal up to a rescaling of $\Delta \rho$.  This suggests that the behavior of the full noise kernel in de Sitter space can tell us something significant about the behavior of the full noise kernel in the near horizon region in Schwarzschild and provides evidence for our conjecture that the validity of the short distance expansion for the two metrics should be similar (at least for separations in the time or radial directions).

In~\cite{EBRAH}, we noted that we expected the quasi-local expansion to be valid when the geodesic distance was small in comparison to the mass scale of the black hole in the region near the horizon.  For de Sitter space, the corresponding distance scale is given by the Hubble parameter $\alpha$. To investigate the range of validity of the expressions above, we compute the relative error between the exact expression for the noise kernel and the truncated series expansions,
\bea
  \left|  \frac{[N_{\hat{T}\hat{T}\hat{T'}\hat{T'}}(x,x')]_{\text{series}}}{ N_{\hat{T}\hat{T}\hat{T'}\hat{T'}}(x,x')} - 1 \right| \;. \label{error}
\eea
For the purposes of our investigation, we define the region of validity to be the region within which the relative error is less than $10\%$.

\begin{figure}
\vskip -0.2in \hskip 0.4in
\includegraphics[width=5in,clip]{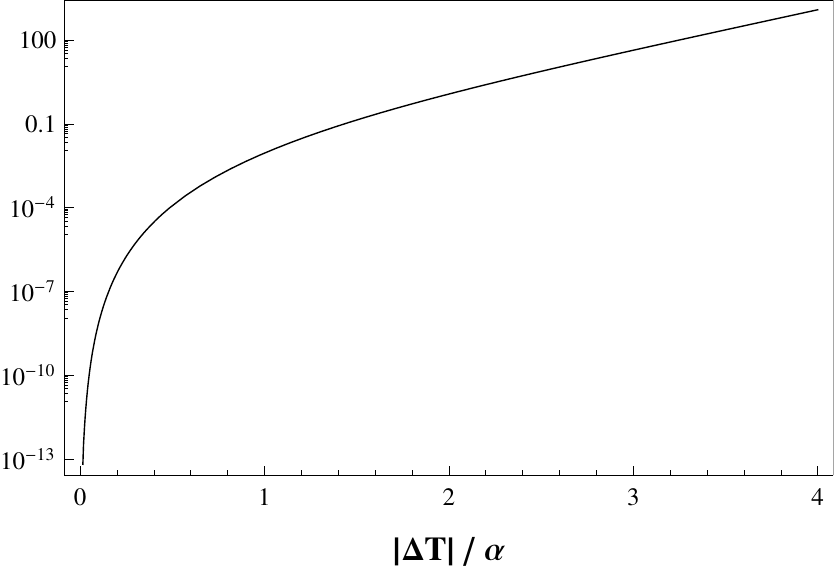}
\vskip 0.2in
\caption{This figure shows the relative error in Eq.~\eqref{error} due to separation in $T$, with $\Delta \rho = \gamma = 0$.  This error is independent of the distance from either point to the cosmological horizon.}
\label{fig41}
\end{figure}

\begin{figure}
\vskip -0.2in \hskip 0.4in
\includegraphics[width=5in,clip]{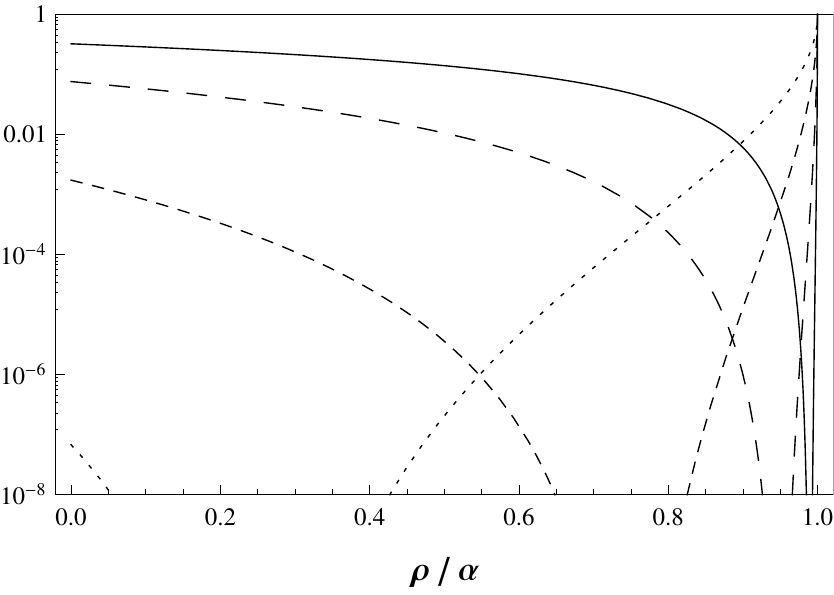}
\vskip 0.2in
\caption{This figure shows the relative error in Eq.~\eqref{error} due to changes in $\rho$ for $\rho'=0.99\alpha$ (solid line), $0.95\alpha$ (long dashes), $0.75\alpha$ (medium dashes), and $0.25\alpha$ (short dashes) with $\Delta T = \gamma = 0$.  On this scale, $\rho/\alpha=1$ marks the cosmological horizon.}
\label{fig42}
\end{figure}

\begin{figure}
\vskip -0.2in \hskip 0.4in
\includegraphics[width=5in,clip]{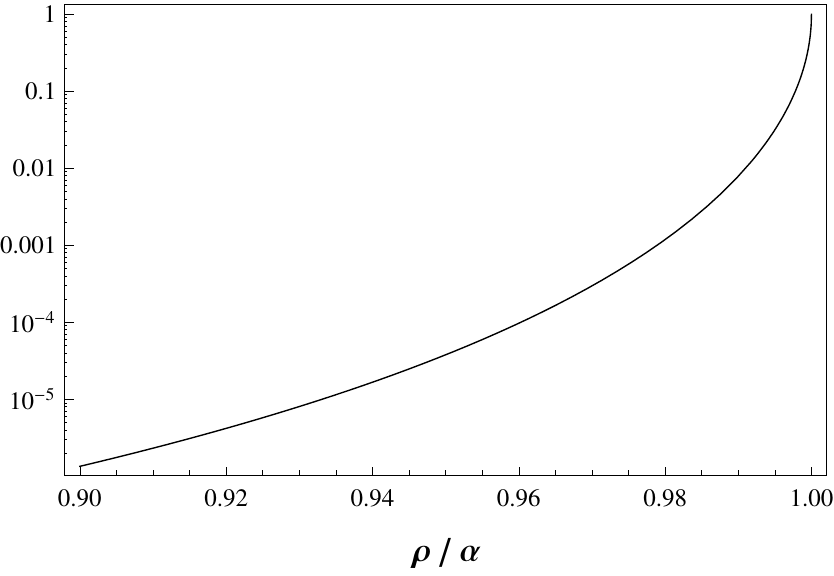}
\vskip 0.2in
\caption{This figure shows the relative error in Eq.~\eqref{error} due to fixed radial separations of $\Delta \rho = 0.1 \alpha$ as both points near the horizon, with $\Delta T = \gamma = 0$.  On this scale, $\rho/\alpha=1$ marks the cosmological horizon.}
\label{fig46}
\end{figure}

When the points are separated in time, we find that the error between the truncated series and the exact expression for static de Sitter space is approximately $10\%$ when $|\Delta T | \approx 1.5 \alpha$ (see Fig.~\ref{fig41}).   This result is independent of the distance between the two points and the horizon, as can be seen by inspection of Eqs.~\eqref{nk-time-ex} and~\eqref{nk-time}.  For radial separation, the situation is somewhat different.  Unlike in Schwarzschild spacetime, where we are interested in points outside the event horizon, in the static de Sitter case we are interested in the region inside the cosmological horizon; thus, the radial separation can never be larger than $\alpha$ without one point crossing the horizon or the origin.  What we find is that the error remains small so long as both points are sufficiently far from the horizon.  As either point nears the cosmological horizon, we find that region of validity scales roughly linearly with the distance from the horizon.  Figures~\ref{fig42} -~\ref{fig46} illustrate this behavior.

In contrast with the cases of time and radial separation, this type of analysis fails when the points are separated in only the angular direction.  The reason for this is that when we expand Eq.~\eqref{NKstatic-TTTT-orth} in powers of $1/\eta$ with $\Delta T = \Delta \rho = 0$, we find that the resulting series contains only terms up to order $O[(x-x')^{-4}]$ and thus provides an exact expression for the noise kernel.  Explicitly, we find
\bea
  [N_{\hat{T}\hat{T}\hat{T'}\hat{T'}}(x,x')]_{\text{series}} &=& -\frac{1}{128 \pi^4 \eta^6 \rho^8} - \frac{1}{64 \pi^4 \eta^5 \rho^8} + \frac{1}{128 \pi^4 \eta^4 \rho^8} \nn
   &=& N_{\hat{T}\hat{T}\hat{T'}\hat{T'}}(x,x') \;. \label{nk-ang}
\eea
However, we suspect that this behavior is a coincidence due to the symmetries present in de Sitter space and that this result will likely not hold for Schwarzschild spacetime.

\section{Discussion}
\label{disc}

We have computed an expression for the noise kernel for the conformally invariant scalar field in any metric that is conformal to the full Minkowski metric when the field is in the conformal vacuum state.  While it could be used for other conformally flat metrics, for metrics which are conformal to Rindler space the Rindler vacuum would be the more natural one to use.  A general form for this expression is shown in Eq.~\eqref{conf-delta} when the points are separated in an arbitrary direction.  For null separations, the noise kernel exhibits a $\delta^2 (\sigma)$ term arising from the square of the Wightman function; as a result, integrating the noise kernel against any function which does not vanish on the light cone will result in a $\delta(0)$ divergence.  The exact meaning of this divergence is under investigation.

Using the above result, we have computed an exact expression for the noise kernel in the conformal vacuum in de Sitter spacetime.  Using a short distance expansion and analytically continuing this expression to the Euclidean sector, we found that Eq.~\eqref{conf-delta} agrees with the expression computed by Osborn and Shore~\cite{o-s-99} for the two point stress tensor correlator on $S^4$.
 We also made a coordinate transformation to the static de Sitter coordinate system and evaluated the resulting expressions for the noise kernel in an orthonormal frame.  All components have been computed for an arbitrary separation of the points.  However, for brevity only two components are displayed in Eq.~\eqref{NKstatic-orth} for the case when the points are spacelike or timelike separated.  The expressions are extremely long if the delta function terms, which only contribute if there is a null separation, are included.

Investigating these expressions in the region near the cosmological horizon, we found that all components of the noise kernel remained finite so long as the points are not null-separated.  However, since $T$ is a null coordinate at the horizon, $\Delta T$ is ill defined when one or both points are on the horizon; thus, we are restricted to cases where the points are arbitrarily close to but not on the horizon.

Since the metric for the static de Sitter coordinates is similar in form to the metric for Schwarzschild spacetime, with the event horizon replaced by the cosmological horizon for observers at the origin, it is expected that the behavior of the noise kernel near the cosmological horizon in the static de Sitter coordinates should be similar to the behavior of the noise kernel near the event horizon in Schwarzschild spacetime where only an approximate solution has been obtained~\cite{EBRAH}.   In the static de Sitter coordinate system, the state we use is equivalent to a Gibbons-Hawking state~\cite{gibbons-hawking} with $\kappa = 1 / 4\alpha$; as a result, the noise kernel for this state may be compared with the noise kernel for Schwarzschild spacetime when the field is in the Hartle-Hawking state~\cite{hartle-hawking}, with the Hubble distance $\alpha$ playing the role of the mass $M$.  Thus, Eq.~\eqref{NKstatic-orth} may be compared directly with the approximate expression for the noise kernel in Schwarzschild spacetime obtained in~\cite{EBRAH}, where an expansion in terms of inverse powers of the geodesic separation was used.  To do this, we obtained a similar expression for the noise kernel for the static de Sitter coordinates by expanding the expression for the $N_{\hat{T}\hat{T}\hat{T'}\hat{T'}}$ component in Eq.~\eqref{NKstatic-TTTT-orth} in powers of $\Delta T$ and $\Delta \rho$.  As expected, we found that the leading order behaviors of both expansions are identical, and that the remaining orders are similar in form, although the coefficients are different.

It is also of interest to investigate the range of validity of the quasi-local approximation used in~\cite{EBRAH}.  To do so, we compared the exact result for the noise kernel computed in the static de Sitter coordinates with the truncated series expansion.  In general, we found that the quasi-local approximation remains valid for time separations smaller than the Hubble distance $\alpha$, and found an error of approximately $10\%$ when $\Delta T \approx 1.5\alpha$.  For radial separations, we are restricted to separations no greater than $\alpha$.  We found that the relative error remains small so long as both points are sufficiently far from the cosmological horizon, but goes to infinity as either point approaches the horizon.  When both points are near the horizon, we find that the region of validity scales roughly linearly with the distance between the point nearest the horizon and the horizon itself.  Finally, for angular separations, we find that the expression generated by the expansion procedure is equal to the exact expression if all three terms are kept; however, we suspect that this result is an artifact of the symmetry of de Sitter space and will not hold in Schwarzschild spacetime.

From these results, we expect the quasi-local approximation used in~\cite{EBRAH} to be valid when the separation of the points is less than the mass scale, so long as neither point is too near the horizon.  In addition, we expect for radial separations that the range of validity should scale roughly linearly with distance from the horizon.  Finally, if one or both points are on the horizon, we expect the quasi-local approximation to be invalid; however, based on our results in static de Sitter space we expect that the exact noise kernel for Schwarzschild spacetime will remain finite on the horizon so long as the two points are not null-separated.

The comparison  made here between the noise kernel in static de Sitter spacetime with respect to the Gibbons-Hawking vacuum and that in the Schwarzschild spacetime with respect to the Hartle-Hawking vacuum, both in the near-horizon region, is useful for a study of the backreaction of Hawking radiation on a quasi-static black holes enclosed in a box and the behavior of the quantum field-induced metric fluctuations via the Einstein-Langevin equation. It is our hope to  explore further along this direction in subsequent papers.

\section*{Acknowledgements}
We would like to thank Albert Roura for  helpful conversations.  This work was supported in part by the National
Science Foundation under Grant No. PHY-0856050 to Wake Forest University and PHY-0801368 to the University of Maryland.  Jason Bates and Hing Tong Cho are supported in part by the National Science Council of the Republic of China under the grants NSC 99-2112-M-032-003-MY3, NSC 101-2811-M-032-005, and by the National Center for Theoretical Sciences (NCTS).


\begin{thebibliography}{100}

\bibitem{bd-book} See e.g. N.~D. Birrell and P.~C.~W. Davies, {\em Quantum fields in curved space} (Cambridge University Press, Cambridge, 1994).

\bibitem{Haw74} S. W. Hawking, Nature (London), {\bf 248}, 30; Comm. Math. Phys. {\bf 43}, 199 (1975).

\bibitem{Guth}  A. H. Guth, 
Phys. Rev. D 23, 347 (1981).

\bibitem{inflation-review} see e.g. A. D. Linde, {\it Particle Physics and Inflationary Cosmology} (Harwood, Chur, Switzerland, 1990).

\bibitem{HuRouVer07} B. L. Hu, A. Roura, and E. Verdaguer,
 Phys. Rev. D {\bf 70}, 044002 (2004).

\bibitem{And-Mol-Mot-1} P. R. Anderson, C. Molina-Paris, and E. Mottola, Phys. Rev. D {\bf 67}, 024026 (2003).


\bibitem{RouVer} A. Roura and E. Verdaguer,
Phys. Rev. D {\bf 78}, 064010 (2008).


\bibitem{PRV} G. P�rez-Nadal, A. Roura, and E. Verdaguer,
Phys. Rev. D {\bf 77}, 124033 (2008).


\bibitem{WNF07} C.-H. Wu, K.-W. Ng, and L.H. Ford,
Phys. Rev. D {\bf 75},  103502 (2007).


\bibitem{FMNWW} L.H. Ford, S.P. Miao, K.-W. Ng, R.P. Woodard, and C.-H. Wu,
Phys. Rev. D {\bf82}, 043501 (2010).


\bibitem{HuRou07} B. L. Hu  and Albert Roura, Phys. Rev. D {\bf 76}, 124018 (2007).

\bibitem{HVCQG} B. L. Hu and E. Verdaguer,  Class. Quant. Grav.  20 (2003)  R1-R42 

\bibitem{stograLivRev} B. L. Hu and E. Verdaguer, Liv. Rev. Rel. {\bf 11}, 3 (2008).

\bibitem{martin99b} R. Mart\'\i n and E. Verdaguer, Phys. Rev. D {\bf 60}, 084008 (1999).

\bibitem{roura99b} A. Roura and E. Verdaguer, Int. J. Theor. Phys. {\bf 38}, 3123 (1999).

\bibitem{PH01} N. G.  Phillips and  B. L. Hu, Phys. Rev. D {\bf 63}, 104001 (2001).

\bibitem{martin99a} R. Mart\'\i n and E. Verdaguer, Phys. Lett. B {\bf 113}, 465 (1999).

\bibitem{martin00} R. Mart\'\i n and E. Verdaguer, Phys. Rev. D {\bf 61}, 124024 (2000).

\bibitem{EBRAH} A. Eftekharzadeh, J. D. Bates, A. Roura, P. R. Anderson, and B. L. Hu, Phys. Rev. D{\bf 85}, 044037 (2012).

\bibitem{phillips03} N. G. Phillips and B. L. Hu, Phys.Rev. D {\bf 67}, 104002 (2003).

\bibitem{roura99} A. Roura and E. Verdaguer, Int. J. Theor. Phys. {\bf 38}, 3123 (1999).

\bibitem{pn09} G. P\`{e}rez-Nadal, A. Roura, and E. Verdaguer, JCAP {\bf 05} 036, (2010).

\bibitem{ChoHu11} H. T. Cho and B. L. Hu, Phys. Rev. D84, 044032 (2011);
      J. Physics (Conf. Ser.) 330, 012002 (2011)  [arXiv:1105.5302]

\bibitem{PH03} N. G. Phillips  and B. L. Hu, Phys. Rev. D {\bf 67}, 104002  (2003).

\bibitem{WHFN} C.-H. Wu, J.-T. Hsiang, L. H. Ford, K.-W. Ng,
Phys. Rev. D {\bf 84}, 103515 (2011).

\bibitem{HRS} B. L. Hu, A. Raval and S. Sinha, ``Notes on Black Hole Fluctuations  and Backreaction", in {\it
    Black Holes, Gravitational Radiation and the Universe: Essays in honor of C. V. Vishveshwara}, edited by B.
    Iyer and B. Bhawal (Kluwer Academic Publishers, Dordrecht, 1998), \eprint{gr-qc/9901010}.

\bibitem{SRH} S. Sinha, A. Raval and B. L. Hu, ``Black Hole Fluctuations and Backreaction in Stochastic Gravity",
    in Foundations of Physics 33 (2003) 37-64  [gr-qc/0210013]

\bibitem {york85} J. W. York, Jr., Phys. Rev. D {\bf 31}, 775 (1985).

\bibitem {York} J. W. York, Jr., Phys. Rev. D {\bf 28}, 2929 (1983); \emph{ibid.} {\bf 33}, 2092 (1986).

\bibitem{York2} D. Hochberg and T. W. Kephart, Phys. Rev. D {\bf 47}, 1465 (1993); D. Hochberg, T. W. Kephart,
    and J. W. York, Jr., Phys. Rev. D {\bf 48}, 479 (1993);  P. R. Anderson, W. A. Hiscock, J. Whitesell, and J.
    W. York, Jr., Phys. Rev. D {\bf 50}, 6427 (1994).

\bibitem{Isham78} S. J. Avis, C. J. Isham, and D. Storey, Phys. Rev. D {\bf 18}, 3565 (1978).

\bibitem{hartle-hawking} J. B. Hartle and S. W. Hawking, Phys. Rev. D {\bf 13}, 2188 (1976); W. Isreal, Phys. Lett. A {\bf 57}, 107 (1976).

\bibitem{o-s-99} H. Osborn and G. M. Shore, Nuc. Phys. B {\bf 571}, 287 (2000).

\bibitem{MTW} C. W. Misner, K. S. Thorne, and J. A. Wheeler, {\it Gravitation} (Freeman, San Francisco, 1973).

\bibitem{a-j-86} B. Allen and T. Jacobson, Commun. Math. Phys. {\bf 103}, 669 (1986).

\bibitem{boulware} D. G. Boulware, Phys. Rev. D {\bf 11}, 1404 (1975); {\bf 13} 2169 (1976).

\bibitem{gibbons-hawking} G. W. Gibbons and S. W. Hawking, Phys. Rev. D {\bf 15}, 2738 (1977).

\end{thebibliography}
\end{document}